\pdfoutput=1
\documentclass[
 aip,
 jcp,
 amsmath,amssymb,
 reprint,
 superscriptaddress,
 floatfix,
]{revtex4-2}

\usepackage{iftex}
\ifpTeX
  \usepackage[dvipdfmx]{graphicx}
  \usepackage[dvipdfmx,dvipsnames]{xcolor}
\else
  \usepackage{graphicx}
  \usepackage[dvipsnames]{xcolor}
\fi
\graphicspath{{figures/}}
\usepackage{bm}
\usepackage{siunitx}
\usepackage{booktabs}
\usepackage{placeins}
\ifpTeX
  \usepackage[dvipdfmx]{hyperref}
\else
  \usepackage{hyperref}
\fi
\hypersetup{colorlinks=true,linkcolor=NavyBlue,citecolor=NavyBlue,urlcolor=NavyBlue}

\begin{document}

\title{Steady shear rheology of a granular crystal containing a single dislocation}
\author{Fumiaki Nakai}
\email{fumiaki.nakai@ess.sci.osaka-u.ac.jp}
\affiliation{Department of Earth and Space Science, Graduate School of Science,
The University of Osaka, 1-1 Machikaneyama, Toyonaka, Osaka 560-0043, Japan}

\begin{abstract}
Monodisperse granular particles can form crystals whose yielding behavior is
strongly affected by dislocations and differs markedly from that of
conventional amorphous granular materials.
Yet the rate dependence of their post-yield
steady rheology remains unclear. We use the discrete element method to study
steady shear in a granular crystal containing a single dislocation. We
find that the steady shear-to-normal stress ratio $\mu_b$ is organized by the
scaled dislocation velocity $v_d/v_s$, rather than by the conventional
inertial number $I$. Here, $v_d$
is related to the imposed shear rate through Orowan kinematics, and $v_s$ is
a characteristic Hertzian elastic-wave speed. At low $v_d/v_s$, the stress
ratio approaches a small plateau associated with the elastic lattice barrier
and interparticle friction. At intermediate values of $v_d/v_s$, contact
damping strongly affects the approximately linear increase of the stress
ratio above the plateau. As $v_d/v_s$ approaches unity, the stress develops a
stronger nonlinear velocity dependence. At still higher velocities, the
coordination deficit rises
sharply, marking the breakdown of crystalline order and the end of the
single-dislocation description. These results identify the scaled dislocation
velocity as the relevant rate variable for the steady rheology of
dislocation-mediated granular flow and clarify the distinct roles of
interparticle friction and contact damping in the low- and intermediate-
velocity regimes, respectively.
\end{abstract}

\maketitle

\section{Introduction}

Monodisperse granular particles can pack into ordered
crystals~\cite{Bragg1947-ua,Karuriya2023-me,Karuriya2024-ea}, whose mechanical
behavior differs markedly from that of amorphous granular
materials~\cite{Andreotti2013-yo,Nicolas2018-zj}. In these athermal solids,
crystalline order and lattice defects strongly affect the macroscopic
mechanical response~\cite{Nakai2026-zo}. Crystalline order raises the yield
stress severalfold above the amorphous
value~\cite{Karuriya2023-me,Karuriya2024-ea}, whereas
introducing a single dislocation lowers it by one to three orders of magnitude
relative to that of a perfect crystal~\cite{Nakai2025-fj,Nakai2025-ef}.
Previous studies of these granular crystals, however, have focused mainly on
the onset of yielding. Their post-yield steady rheology, particularly the
dependence of steady stress on the imposed shear rate, remains largely
unexplored.

\begin{figure}[!t]
    \centering
    \includegraphics[width=1.0\linewidth]{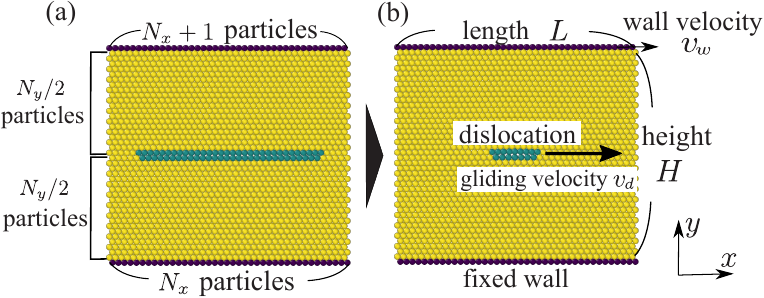}
    \caption{System setup.
    (a) Initial configuration with a mismatched interface (teal). Particles
    are arranged on triangular lattices with nearest-neighbor spacing
    $b=(1-\alpha)d$, where $\alpha$ is the initial overlap ratio. The system
    length and wall separation are $L=N_xb$ and
    $H=\sqrt{3}(N_y+1)b/2$, respectively. The upper and lower halves contain
    $(N_x+1)N_y/2$ and $N_xN_y/2$ mobile particles, respectively. The extra
    column in the upper half is accommodated within the same system length
    $L$, thereby seeding a lattice mismatch.
    (b) Configuration obtained after integrating the equations of motion for
    $10^6$ time steps. The initial mismatch relaxes into a crystal containing
    a single dislocation. The top wall is subsequently driven at
    $v_w=\dot{\gamma}H$, while the bottom wall remains fixed. When the imposed
    strain is carried by this dislocation, Orowan kinematics gives
    $v_d=LH\dot{\gamma}/b$, as verified previously for this
    geometry~\cite{Nakai2025-fj,Nakai2025-ef}.
    }
    \label{fig:setup}
\end{figure}

We investigate this post-yield rheology using the model granular crystal
containing a single edge dislocation that we studied
previously~\cite{Nakai2025-fj,Nakai2025-ef}. That work established the onset
and kinematics of dislocation glide, including the associated yield stress,
for this geometry. As in atomic crystals, plastic shear is carried by the
dislocation, and Orowan kinematics specifies $v_d$ for a given imposed shear
rate and geometry~\cite{Anderson2017-ry}. This kinematic relation does not,
however, determine how the steady rheology depends on $v_d$. In atomic
crystals, it must be supplemented by a velocity--stress mobility
law~\cite{Anderson2017-ry,Joos2001-au,Olmsted2005-ls,Fan2021-fs}, whereas no
corresponding relation has been established for the present granular crystal.
From the viewpoint of athermal granular rheology, the standard description of
dense flow is the $\mu(I)$ rheology, in which the shear-to-normal stress ratio is expressed
as a function of the inertial number $I$. Thus, $I$ is a plausible candidate
for the relevant rate variable in the present
system~\cite{GDR-MiDi2004-ay,Pouliquen1999-ca,Jop2006-yn,Forterre2008-cr,Andreotti2013-yo,Kamrin2024-ig}.
It is not known a priori whether this conventional description applies when
plastic strain is carried by a single dislocation. Moreover, the effects of
interparticle friction and dissipative contact damping on the resulting
steady rheology remain unclear.

We therefore perform simulations using the discrete element method (DEM) to
examine the post-yield steady rheology of a granular crystal containing a
single dislocation, varying the initial precompression, system size, contact
damping, and interparticle friction. We find that the steady shear-to-normal
stress ratio, a natural stress measure in granular rheology, is organized more
effectively by the scaled dislocation velocity $v_d/v_s$ than by the
conventional inertial number $I$. The response exhibits four velocity regimes.
At small $v_d/v_s$, the stress ratio approaches a small plateau associated
with the elastic lattice
barrier~\cite{Peierls1940-ah,Nabarro1947-cj,Joos1997-zn,Edagawa2019-iw}, and
interparticle friction raises this plateau. At intermediate $v_d/v_s$, contact
damping strongly affects the approximately linear increase above the plateau.
As $v_d/v_s$ approaches unity, the stress develops a stronger nonlinear
velocity dependence. At higher imposed rates, the coordination deficit rises
sharply, marking the breakdown of crystalline order and the limit of the
single-dislocation description. These results identify $v_d/v_s$ as an
effective rate variable for the steady stress of dislocation-mediated
granular flow and distinguish the effects of interparticle friction and
contact damping in the low- and intermediate-velocity regimes, respectively.

\begin{table}[htbp]
\centering
\caption{Parameters for the DEM simulations using the Hertz--Mindlin--Tsuji
contact model. The material constants are based on a typical
elastomer~\cite{callister2020materials,Robertson2007-sk}. All combinations of
the listed system sizes, $\alpha$, $\dot\gamma$, $\mu_p$, and $e$ were
simulated, giving 4992 runs.}
\label{table:parameters}
\scriptsize
\resizebox{\columnwidth}{!}{%
\begin{tabular}{@{}lcc@{}}
\toprule
Parameter & Symbol & Value \\ \midrule
Particle mass density         & $\rho$         & \SI{1000}{\kilogram\per\cubic\metre} \\
Particle diameter             & $d$            & \SI{1}{\milli\metre} \\
Young's modulus               & $E$            & \SI{1}{\mega\pascal} \\
Poisson's ratio               & $\nu$          & $0.45$ \\
Initial overlap (precompression) ratio
                               & $\alpha$       & $0.03$, $0.05$, $0.07$, $0.09$ \\
System size                   & $(N_x,N_y)$     & $(100,40)$, $(200,40)$, $(400,40)$, $(100,80)$ \\
Number of mobile particles    & $N$            & $4020$, $8020$, $16020$, $8040$, respectively \\
Periodic system length        & $L$            & $N_x(1-\alpha)d$ \\
Wall separation               & $H$            & $\sqrt{3}(N_y+1)(1-\alpha)d/2$ \\
Shear rate                    & $\dot{\gamma}$ & \shortstack{$0.003$, $0.005$, $0.01$, $0.02$, $0.03$, $0.05$,\\
                                                              $0.1$, $0.2$, $0.3$, $0.5$, $1$, $2$, $3\,\si{\per\second}$} \\
Interparticle friction coefficient
                               & $\mu_p$        & $0$, $10^{-4}$, $10^{-3}$, $10^{-2}$, $0.1$, $0.3$ \\
Restitution coefficient       & $e$            & $0.2$, $0.4$, $0.6$, $0.8$ \\
\bottomrule
\end{tabular}
}
\end{table}

\FloatBarrier

\section{Simulation Method}

\subsection{Contact model and simulation parameters}

We perform simulations using the discrete element method (DEM) as implemented
in LAMMPS~\cite{Thompson2022-lammps}, following the setup of our previous
work~\cite{Nakai2025-fj,Nakai2025-ef}. The particles are three-dimensional
spheres whose translational motion is constrained to the $xy$ plane, while
rotation is allowed about the $z$ axis. The system is periodic in $x$ and
confined between particle walls in $y$. Each sphere has diameter
$d=\SI{1}{\mm}$, mass density $\rho=\SI{1000}{\kg\per\cubic\m}$, and mass
$m=\pi\rho d^3/6$. When converting the calculated virial to stress, we assign
the monolayer an effective thickness $d$. The Young's modulus $E$ and
Poisson's ratio $\nu$ are chosen to represent a typical elastomer and are
listed in Table~\ref{table:parameters}.

The motion of each mobile particle is governed by the translational and
rotational equations under the Hertz--Mindlin--Tsuji contact model:
\begin{align}
    m\ddot{\bm{r}}_{i}
    &=\sum_{j\ne i}\bm{F}_{ij}
    =\sum_{j\ne i}\left(F_{n,ij}\bm{n}_{ij}+\bm{F}_{t,ij}\right)
    \Theta(d-r_{ij}), \\
    \mathcal I\dot{\bm{\omega}}_i
    &=\sum_{j\ne i}\left(-\frac{d-\xi_{n,ij}}{2}\bm{n}_{ij}\right)
    \times\bm{F}_{t,ij}\,\Theta(d-r_{ij}),
\end{align}
Here, the sums include contacts with both mobile and wall particles.
The quantities $\bm r_i$ and $\bm\omega_i$ denote the position and angular
velocity of particle $i$, and $\mathcal I=md^2/10$ is the moment of inertia of
a uniform sphere. The relative position vector is
$\bm r_{ij}\equiv\bm r_i-\bm r_j$, with $r_{ij}=|\bm r_{ij}|$ and
$\bm n_{ij}=\bm r_{ij}/r_{ij}$. Thus, $\bm n_{ij}$ points from particle $j$ to
particle $i$, and $\bm F_{ij}$ is the force exerted on $i$ by $j$. The factor
$\Theta(d-r_{ij})$ restricts the interaction to overlapping particles.

The normal contact force consists of an elastic Hertzian repulsion and a
velocity-dependent damping term,
\begin{align}
    F_{n,ij}&=k_n\xi_{n,ij}^{3/2}+\eta_{ij}v_{n,ij},
    \label{eq:normal_force}\\
    v_{n,ij}&=-\bm{v}_{ij}\cdot\bm{n}_{ij},\\
    \bm{v}_{t,ij}&=\bm{v}_{ij}
    -(\bm{v}_{ij}\cdot\bm{n}_{ij})\bm{n}_{ij}
    -\frac{d}{2}(\bm{\omega}_i+\bm{\omega}_j)\times\bm{n}_{ij},
\end{align}
where $\bm v_{ij}=\dot{\bm r}_i-\dot{\bm r}_j$ and
$\xi_{n,ij}=d-r_{ij}$. The uncapped tangential force is
\begin{equation}
    \bm F^{*}_{t,ij}=-k_t\sqrt{\xi_{n,ij}}\,\bm\xi_{t,ij}
    -\eta_{ij}\bm v_{t,ij}.
\end{equation}
Here, $\bm\xi_{t,ij}$ is the accumulated tangential displacement. The
Coulomb cap is imposed as
\begin{equation}
    \bm F_{t,ij}=\min\!\left(
    1,\frac{\mu_p|F_{n,ij}|}{|\bm F^{*}_{t,ij}|}
    \right)\bm F^{*}_{t,ij}.
    \label{eqmain:force_tangential}
\end{equation}
Here, $\mu_p$ is the interparticle friction coefficient. At each time step,
$\bm\xi_{t,ij}$ is rotated to remain perpendicular to
the current contact normal $\bm n_{ij}$, following the LAMMPS
implementation~\cite{Luding2008-su,Thompson2022-lammps}.

For two identical spheres, the Hertz--Mindlin stiffnesses reduce to
\begin{equation}
    k_n=\frac{E\sqrt{d}}{3(1-\nu^2)}, \qquad
    k_t=\frac{E\sqrt{d}}{(1+\nu)(2-\nu)} .
\end{equation}
We set the tangential-to-normal damping ratio to unity. The resulting normal
and tangential damping coefficients are therefore both
\begin{equation}
    \eta_{ij}=\kappa\sqrt{\frac{m k_n\sqrt{\xi_{n,ij}}}{2}},
    \label{eq:damping_coefficient}
\end{equation}
where the Tsuji damping factor is related to the input restitution coefficient
$e$ by~\cite{Tsuji1992-dem}
\begin{equation}
\begin{split}
    \kappa={}&1.2728-4.2783e+11.087e^2-22.348e^3\\
             &+27.467e^4-18.022e^5+4.8218e^6.
\end{split}
\label{eq:kappa}
\end{equation}
The integration time step is
$\Delta t=\SI{8.9e-6}{\second}$.

\subsection{System preparation and shear protocol}
\label{sec:shear_protocol}

We construct an initial configuration designed to relax into a single edge
dislocation by joining two triangular crystals with slightly different column
spacings [Fig.~\ref{fig:setup}(a)]. The two halves occupy the same simulation
cell of width $L$, with periodic boundary conditions along $x$. In the lower
half, the $N_x$ wall particles and $N_x\times N_y/2$ mobile particles are
placed on a triangular lattice with lattice vectors $((1-\alpha)d,\,0)$ and
$\left((1-\alpha)d/2,\,\sqrt{3}(1-\alpha)d/2\right)$. Here,
$\alpha\equiv(d-b)/d$ is the initial overlap ratio and
$b=(1-\alpha)d$ is the lattice spacing, which also gives the magnitude of the
Burgers vector of the resulting dislocation. In the upper half, the $N_x+1$
wall particles and $(N_x+1)\times N_y/2$ mobile particles are placed with the
horizontal repeat distance reduced by the factor $N_x/(N_x+1)$. The
corresponding lattice vectors are
$\left(N_x(1-\alpha)d/(N_x+1),\,0\right)$ and
$\left((1-\alpha)d/2,\,\sqrt{3}(1-\alpha)d/2\right)$.
The upper crystal therefore contains one additional particle column within
the same width $L$, creating the mismatch that seeds the dislocation.

We relax this initial structure for $10^6$ time steps at $\mu_p=0$, using the
damping corresponding to the same restitution coefficient as in the
subsequent shear stage. During relaxation, the lower wall is fixed, while the
upper wall is held fixed in $y$ and its particles receive the same
group-averaged tangential force, preserving their relative spacing. The
mismatch thereby relaxes into the single edge dislocation shown in
Fig.~\ref{fig:setup}(b), without prescribing its core structure. This
construction and the resulting glide kinematics were verified in our previous
work~\cite{Nakai2025-fj,Nakai2025-ef}.

After relaxation, we set the target interparticle friction coefficient
$\mu_p$ and shear the system by moving the top wall at
$v_w=\dot{\gamma}H$ while keeping the bottom wall fixed. The imposed strain is
$\gamma=\dot{\gamma}t$. All production runs use the same number of shear
steps, $11235956$. The final strain therefore ranges from approximately
$0.3$ at $\dot\gamma=\SI{0.003}{\per\second}$ to $300$ at
$\dot\gamma=\SI{3}{\per\second}$. For an observable $X$, we define its steady
value $\bar X$ as the average over the final 20\% of the recorded data. The
same trends are obtained when the averaging window is extended to the final
50\%.

\subsection{Measured quantities and characteristic scales}
\label{sec:scales}

Before presenting the results, we summarize the principal physical quantities
and dimensionless variables used in this study. We use the following
per-particle stress convention:
\begin{equation}
    S_{ab}^{(i)}=-m_i v_{i,a}v_{i,b}
    -\frac{1}{2}\sum_{j\ne i}r_{ij,a}F_{ij,b}.
    \label{eq:atom_stress}
\end{equation}
This expression includes the raw-velocity kinetic term and the virial from
contacts with both mobile and wall particles. The stress components are
defined as
\begin{equation}
    \sigma_{ab}=-\frac{1}{V}\sum_{i\in{\rm mobile}}S_{ab}^{(i)},
    \qquad ab\in\{xx,yy,xy\},
    \label{eq:stress}
\end{equation}
so that compressive normal stresses are positive
~\cite{Evans-J2007-ft,Allen2017-ja}. The quasi-two-dimensional volume is
$V=LHd$.

We define the steady stress ratio
$\mu_b=-\bar\sigma_{xy}/\bar\sigma_{yy}$ and the inertial number
$I=\dot{\gamma}d/\sqrt{\bar{\sigma}_{yy}/\rho}$
~\cite{Andreotti2013-yo}. We denote by $Z$ the contact number averaged over
the $N$ mobile particles, including contacts with both mobile and wall
particles. The quantity $N(6-Z)$ provides an extensive measure of the
coordination deficit relative to a perfect sixfold crystal. It is
approximately independent of system size for a single localized dislocation
and increases when the crystalline contact network breaks down.

When the imposed plastic strain is carried by a single dislocation, Orowan
kinematics gives~\cite{Nakai2025-fj,Nakai2025-ef}
\begin{equation}
    v_d=\frac{LH\dot{\gamma}}{b}.
    \label{eq:orowan}
\end{equation}
Within the single-dislocation-glide regime, $v_d$ represents the time-averaged
glide velocity of the dislocation. This interpretation no longer applies once
the crystalline structure breaks down.

For Hertzian contacts, the initial overlap ratio $\alpha$ controls the
precompression and sets the normal-stress scale
$\bar{\sigma}_{yy}\sim E\alpha^{3/2}$. Linearizing the normal force about the
overlap $\xi_n\sim\alpha d$ gives an effective stiffness
$k_{\rm eff}\sim Ed\sqrt{\alpha}$ and hence the characteristic elastic-wave
speed
\begin{equation}
    v_s\equiv\sqrt{\frac{E\sqrt{\alpha}}{\rho}}.
    \label{eq:sound}
\end{equation}
We use $v_s$ only as a characteristic sound-speed scale and do not distinguish
between longitudinal and transverse modes~\cite{Velicky2002-vm}.

\subsection{Use of artificial intelligence tools}

OpenAI Codex (GPT-5, OpenAI; accessed May--August 2026) was used
interactively to assist in developing and debugging simulation input files
and data-analysis and plotting scripts, and in editing the English text, with
the aims of improving development efficiency and identifying implementation,
consistency, and language errors.
The author performed and verified all simulations and quantitative analyses,
checked the reported data and figures, and made all scientific judgments and
interpretations.

\section{Results}

\subsection{Stress--strain response}

\begin{figure}[!t]
    \centering
    \includegraphics[width=0.95\linewidth]{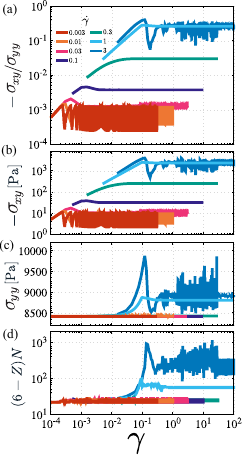}
    \caption{Stress--strain response of the single-dislocation crystal for
    $N_x=400$, $N_y=40$, $\alpha=0.05$, $\mu_p=0$, and $e=0.4$.
    Colors indicate the imposed shear rates
    $\dot\gamma=0.003$, $0.01$, $0.03$, $0.1$, $0.3$, $1$, and
    $\SI{3}{\per\second}$. The panels show (a) the instantaneous stress
    ratio $-\sigma_{xy}/\sigma_{yy}$, (b) the shear stress $-\sigma_{xy}$,
    (c) the normal stress $\sigma_{yy}$, and (d) the extensive coordination
    deficit $N(6-Z)$. At the lowest rates, the stress ratio is of order
    $10^{-3}$, approximately two orders of magnitude below the typical
    order-$10^{-1}$ value of dense amorphous granular flows. The inertial number
    remains in the range $I\simeq10^{-6}$--$10^{-3}\ll1$, where dense amorphous
    granular flows are nearly shear-rate independent~\cite{GDR-MiDi2004-ay,Andreotti2013-yo}, yet the stress ratio in
    the present crystal varies strongly with the shear rate. At higher rates, the increase in
    $N(6-Z)$ signals the breakdown of crystalline order.}
    \label{fig:ss}
\end{figure}

Figure~\ref{fig:ss} shows representative stress--strain responses of the
frictionless single-dislocation crystal for $N_x=400$, $N_y=40$,
$\alpha=0.05$, and $e=0.4$. The imposed shear rate ranges from
$\dot\gamma=\SI{0.003}{\per\second}$ to $\SI{3}{\per\second}$, spanning
three decades. At the lowest rates, the stress ratio
$-\sigma_{xy}/\sigma_{yy}$ remains of order $10^{-3}$
[Fig.~\ref{fig:ss}(a)]. This is approximately two orders of
magnitude below the typical $0.1$--$0.4$ range of dense amorphous granular
flows~\cite{Pouliquen1999-ca,GDR-MiDi2004-ay,Jop2006-yn}.

Another notable feature is the strong rate dependence of the response. As
$\dot\gamma$ increases, both the shear stress $-\sigma_{xy}$ and the stress
ratio $-\sigma_{xy}/\sigma_{yy}$ increase systematically. At the largest
rates, the stress ratio reaches order $10^{-1}$ [Fig.~\ref{fig:ss}(a)].
The normal stress remains of order $10^4\,\si{\pascal}$
[Fig.~\ref{fig:ss}(c)]. The corresponding inertial number, defined using
$\bar\sigma_{yy}$, spans only approximately $10^{-6}$--$10^{-3}$. All cases
therefore satisfy $I\ll1$, including those with a pronounced increase in
stress. In conventional dense granular flows, the stress ratio is nearly
rate independent in this small-$I$
regime~\cite{GDR-MiDi2004-ay,Jop2006-yn,Da_Cruz2005-lr}. The unusually small
low-rate resistance and the pronounced rate dependence at $I\ll1$ therefore
distinguish the present crystal from conventional amorphous granular
materials.

At the highest shear rates, the coordination deficit increases
[Fig.~\ref{fig:ss}(d)], indicating the onset of crystalline-order breakdown.
We next examine the steady rheology and the accompanying structural changes
in detail.

\subsection{Shear-rate dependence}

\begin{figure*}[!t]
    \centering
    \includegraphics[width=0.9\linewidth]{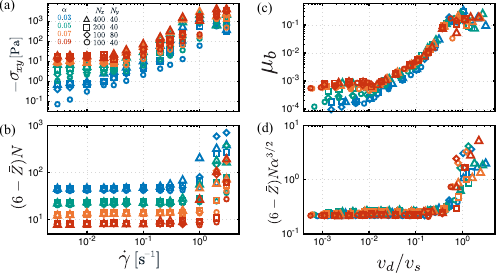}
    \caption{Steady response of the frictionless crystal ($\mu_p=0$ and
    $e=0.4$). Colors denote $\alpha=0.03, 0.05, 0.07, 0.09$, and open-symbol
    shapes distinguish $(N_x,N_y)=(100,40), (200,40), (100,80), (400,40)$,
    as listed in the legend. All 13 imposed shear rates from
    $\dot\gamma=0.003$ to $\SI{3}{\per\second}$ are shown; each point is
    averaged over the final 20\% of the stored time series.
    (a) Steady shear stress $-\bar\sigma_{xy}$ against $\dot\gamma$,
    which increases with $\alpha$ and generally increases with system size.
    (b) Extensive coordination deficit $N(6-Z)$ against $\dot\gamma$. Its
    low-rate plateau corresponds to the single-dislocation crystal, whereas
    its increase at higher rates marks the breakdown of crystalline order.
    The larger plateau at smaller $\alpha$ reflects the spatially extended
    dislocation.
    (c) Stress ratio $\mu_b=-\bar\sigma_{xy}/\bar\sigma_{yy}$ against the
    scaled dislocation velocity $v_d/v_s$. The data collapse reasonably well
    except at $\alpha=0.03$, where the dislocation is spatially extended.
    (d) Scaled extensive coordination deficit
    $N(6-Z)\alpha^{3/2}$ against $v_d/v_s$, where the factor
    $\alpha^{3/2}$ is chosen empirically. Its rise near $v_d/v_s\sim1$
    marks the breakdown of crystalline order as the dislocation velocity
    approaches the characteristic sound-speed scale.}
    \label{fig:stress_rate}
\end{figure*}

We first examine the shear-rate dependence of the steady response while
varying the lattice spacing and system size. The initial overlap $\alpha$
sets the lattice spacing $b=(1-\alpha)d$, an important structural parameter
of the crystal, whereas the system geometry determines the dislocation
velocity required at a given shear rate through the Orowan relation,
$v_d=LH\dot\gamma/b$ [Eq.~\eqref{eq:orowan}]. We vary
$\alpha\in\{0.03,0.05,0.07,0.09\}$ and use
$(N_x,N_y)=(100,40),(200,40),(400,40),(100,80)$, while fixing $\mu_p=0$
and $e=0.4$. All steady values are obtained using the averaging procedure
described in Sec.~\ref{sec:shear_protocol}. In Fig.~\ref{fig:stress_rate},
colors distinguish $\alpha$, and the open-symbol shapes distinguish the
system geometries, as indicated in the legend.

Figure~\ref{fig:stress_rate}(a) shows the steady shear stress
$-\bar\sigma_{xy}$ as a function of $\dot\gamma$ over
$\dot\gamma=\SI{0.003}{\per\second}$--$\SI{3}{\per\second}$. The shear
stress increases with $\dot\gamma$. At the same shear rate,
$-\bar\sigma_{xy}$ is generally larger at higher $\alpha$ and for larger
systems.

Figure~\ref{fig:stress_rate}(b) shows the corresponding extensive
coordination deficit $N(6-Z)$. This quantity vanishes in a perfect triangular
crystal and is small but nonzero in the present crystal containing one
dislocation. For every $\alpha$ and system geometry, $N(6-Z)$ remains
approximately constant at small $\dot\gamma$ and increases at large
$\dot\gamma$. The larger low-rate plateau at smaller $\alpha$ reflects the
spatially extended dislocation, whereas the high-rate increase indicates the
breakdown of the crystalline structure.

We next ask which physical scales characterize the stress--rate relation.
As discussed in Sec.~\ref{sec:scales}, the initial overlap $\alpha$ controls
the confining normal stress, $\bar\sigma_{yy}\sim E\alpha^{3/2}$. Because the
normal stress provides a natural stress scale in granular rheology, we use
the stress ratio $\mu_b=-\bar\sigma_{xy}/\bar\sigma_{yy}$. Through the
Orowan relation [Eq.~\eqref{eq:orowan}], the imposed shear rate determines
the dislocation velocity $v_d$, which is much larger than the wall speed.
One possible internal velocity scale is the characteristic sound-speed scale
$v_s$ [Eq.~\eqref{eq:sound}]. The ratio of dislocation to elastic-wave speed
has also been considered in atomistic crystals
~\cite{Olmsted2005-ls,Fan2021-fs}. We therefore examine $v_d/v_s$ as a
dimensionless rate variable. The conventional $\mu$--$I$ representation,
shown in Fig.~\ref{fig:mu-I} in the Appendix for reference, does not collapse
the data.

Figure~\ref{fig:stress_rate}(c) plots the stress ratio $\mu_b$ against
$v_d/v_s$. Except at $\alpha=0.03$, where the dislocation is spatially
extended, the data for different $\alpha$ and system sizes collapse
reasonably well. At small $v_d/v_s$, $\mu_b$ approaches a plateau in the
$10^{-4}$--$10^{-3}$ range, far below the typical order-$10^{-1}$ stress
ratio of dense amorphous granular flows
~\cite{GDR-MiDi2004-ay,Andreotti2013-yo}. It increases at intermediate
$v_d/v_s$ and ultimately reaches a high-velocity plateau of order $10^{-1}$,
comparable to values in dense amorphous granular flows.

Figure~\ref{fig:stress_rate}(d) shows $N(6-Z)\alpha^{3/2}$ against
$v_d/v_s$. The factor $\alpha^{3/2}$ is introduced empirically to reduce the
low-velocity variation with $\alpha$; the larger deficit at small $\alpha$
reflects the spatially extended dislocation. At small $v_d/v_s$, the scaled
deficit remains small and nearly constant, indicating steady
single-dislocation glide. It begins to increase as $v_d/v_s$ approaches
unity, marking the onset of crystalline-structure breakdown. This increase
approximately coincides with the crossover of $\mu_b$ from its increasing
branch to the high-velocity plateau [Fig.~\ref{fig:stress_rate}(c)]. Below
this crossover, deformation is carried by single-dislocation glide; beyond
it, the single-dislocation description is no longer applicable.

Thus, while the single-dislocation crystal is retained, the steady stress
can be reasonably expressed through the stress ratio $\mu_b$ as a function
of the scaled dislocation velocity $v_d/v_s$. We next examine the effects of
two contact properties characteristic of granular materials: interparticle
friction, characterized by $\mu_p$, and contact damping, quantified by
$\kappa$ in Eq.~\eqref{eq:kappa}.

\subsection{Effect of interparticle friction}
\label{sec:friction}

\begin{figure}[!t]
    \centering
    \includegraphics[width=0.9\linewidth]{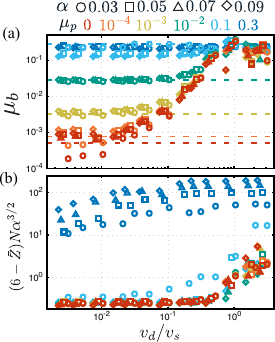}
    \caption{
    Effect of interparticle friction for $N_x=400$, $N_y=40$, and $e=0.4$.
    Open-symbol shapes distinguish $\alpha=0.03, 0.05, 0.07, 0.09$, and colors
    distinguish $\mu_p=0, 10^{-4}, 10^{-3}, 10^{-2}, 0.1, 0.3$, as listed
    above the panels. All 13 imposed shear rates from $\dot\gamma=0.003$ to
    $\SI{3}{\per\second}$ are shown.
    (a) Stress ratio $\mu_b=-\bar\sigma_{xy}/\bar\sigma_{yy}$ against
    $v_d/v_s$. Within the single-dislocation regime, the data for different
    $\alpha$ collapse reasonably well at each $\mu_p$. Interparticle friction
    primarily raises $\mu_b$ at small $v_d/v_s$. The horizontal dashed lines
    show $\mu_b=\mu_0+c\mu_p$, fitted to the low-rate data within the
    single-dislocation regime, with $\mu_0=8.4\times10^{-4}$ and $c=2.8$.
    (b) Scaled extensive coordination deficit
    $N(6-Z)\alpha^{3/2}$ against $v_d/v_s$. Large values occur at high shear
    rates and, for $\mu_p=0.3$, even at low rates, indicating the breakdown
    of the crystalline structure.}
    \label{fig:stress_rate_mu}
\end{figure}

We examine the effect of interparticle friction on the steady rheology.
Figure~\ref{fig:stress_rate_mu}(a) shows $\mu_b$ as a function of $v_d/v_s$
for $N_x=400$, $N_y=40$, and $e=0.4$. Colors indicate $\mu_p$, and symbol
shapes indicate $\alpha$. Within the single-dislocation regime, the data
collapse reasonably well across $\alpha$ at fixed $\mu_p$. Interparticle friction primarily affects the
slow-deformation regime at small $v_d/v_s$, where increasing $\mu_p$ raises
the low-velocity plateau of the stress ratio.

Figure~\ref{fig:stress_rate_mu}(b) shows the scaled coordination deficit
$N(6-Z)\alpha^{3/2}$ against $v_d/v_s$ for the same parameter sets. For
$\mu_p<0.1$, the coordination deficit departs from its low-velocity plateau
as $v_d$ approaches $v_s$, indicating the breakdown of the crystalline
structure. At $\mu_p=0.3$, by contrast, the coordination deficit is already
large at small $v_d/v_s$. Dislocation glide requires relative displacements
between neighboring particles, and strong interparticle friction restricts
this motion. The resulting suppression of glide and breakdown of the
crystalline structure are consistent with our previous findings
~\cite{Nakai2025-fj,Nakai2025-ef}.

Within the low-velocity single-dislocation regime, the stress-ratio plateau
is reasonably described by $\mu_b=\mu_0+c\mu_p$, with
$\mu_0=8.4\times10^{-4}$ and $c=2.8$
[Fig.~\ref{fig:stress_rate_mu}(a)]. Here, $\mu_0$ gives the frictionless
plateau, while $c\mu_p$ gives its leading increase with $\mu_p$. The increase of the
low-velocity plateau with $\mu_p$ is qualitatively similar to that observed
in dense disordered granular flows~\cite{Da_Cruz2005-lr}. In such flows,
however, the bulk stress ratio remains within a relatively narrow range of
order $10^{-1}$, whereas the low-velocity plateau in the present
single-dislocation regime increases by more than an order of magnitude with
$\mu_p$. We discuss the
physical origin of this friction dependence in the Discussion.

\subsection{Effect of contact damping}

\begin{figure}[!t]
    \centering
    \includegraphics[width=0.9\linewidth]{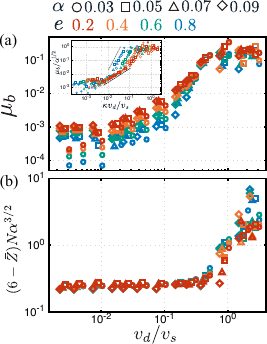}
    \caption{
    Effect of contact damping for $N_x=400$, $N_y=40$, and $\mu_p=0$.
    Colors denote $e=0.2, 0.4, 0.6, 0.8$, and open circles, squares, triangles,
    and diamonds denote $\alpha=0.03, 0.05, 0.07, 0.09$, respectively.
    All 13 imposed shear rates from $\dot\gamma=0.003$ to
    $\SI{3}{\per\second}$ are shown.
    (a) Stress ratio $\mu_b$ against $v_d/v_s$. Except at $\alpha=0.03$, where
    the dislocation is spatially extended, contact damping primarily affects
    the rising regime at intermediate $v_d/v_s$; smaller $e$, corresponding
    to stronger damping, gives larger $\mu_b$. The inset plots
    $\mu_b/\alpha^{1/2}$ against $\kappa v_d/v_s$, where $\kappa(e)$ is given
    by Eq.~\eqref{eq:kappa}. In this representation, the data collapse
    reasonably well in the intermediate-velocity regime. The black and gray
    straight segments have slopes 1 and 3 on the log--log axes, respectively,
    and are guides rather than fits.
    (b) Scaled extensive coordination deficit
    $N(6-Z)\alpha^{3/2}$ against $v_d/v_s$. It increases as $v_d$ approaches
    $v_s$, marking the breakdown of crystalline order, and depends only
    weakly on $e$ over the range studied.}
    \label{fig:damping}
\end{figure}

We next examine contact damping, whose strength is controlled by $e$ through
the damping factor $\kappa$ [Eq.~\eqref{eq:kappa}].
Figure~\ref{fig:damping}(a) plots $\mu_b$ against $v_d/v_s$ for four values
each of $e$ and $\alpha$. Colors distinguish $e$, and open-symbol shapes
distinguish $\alpha$. Except at $\alpha=0.03$, where the dislocation core is
spatially extended, the data for different $\alpha$ collapse reasonably well
at each $e$. The effect of $e$ is most evident on the rising branch at
intermediate $v_d/v_s$: smaller $e$, corresponding to stronger damping,
produces larger $\mu_b$. This sensitivity contrasts with dense disordered
granular flows, in which the restitution coefficient has been reported to
have only a weak effect on the bulk stress ratio~\cite{Da_Cruz2005-lr}.
A weaker dependence on $e$ persists at low $v_d/v_s$, although its origin
remains unclear.

Because $\kappa$ multiplies the relative contact velocity in the damping
force, we examine whether the damping dependence can be represented using
$\kappa v_d/v_s$. The inset plots the empirically scaled stress ratio
$\mu_b/\alpha^{1/2}$ against $\kappa v_d/v_s$. Except for $\alpha=0.03$, the data for different
$e$ and $\alpha$ collapse reasonably well in the intermediate-velocity regime
and show an approximately linear increase, corresponding in this regime to
$\mu_b\sim\kappa\alpha^{1/2}v_d/v_s$. A possible origin of this scaling based
on the Tsuji contact-damping model is discussed in
Sec.~\ref{sec:explanation}. In this rescaled representation,
the nonlinear upturn appears at smaller $\kappa v_d/v_s$ for larger $e$,
corresponding to smaller $\kappa$. The straight segments
with slopes 1 and 3 on the log--log axes are included only as guides to the
eye.

Figure~\ref{fig:damping}(b) plots $N(6-Z)\alpha^{3/2}$ against $v_d/v_s$ for
the same parameter sets. The scaled coordination deficit remains nearly
constant at low velocity and increases as $v_d$ approaches $v_s$, marking
the breakdown of crystalline order. The onset of this increase depends
slightly on $e$, but the shift is weak over the range examined.

\subsection{Empirical representation}

We now combine the preceding observations into an empirical representation
of the steady response in the single-dislocation regime.
Figure~\ref{fig:stress_rate} shows that the frictionless data collapse
reasonably well when expressed in terms of the stress ratio $\mu_b$ and the
scaled dislocation velocity $v_d/v_s$. Figure~\ref{fig:stress_rate_mu} shows
that interparticle friction raises the low-velocity plateau approximately in
proportion to $\mu_p$ while single-dislocation glide persists.
Figure~\ref{fig:damping} suggests an approximately linear contribution
proportional to $\kappa\alpha^{1/2}v_d/v_s$ on the rising branch. At larger
$v_d/v_s$, the nonlinear increase is represented empirically by a cubic term.
Based on these observations, we write
\begin{equation}
    \mu_b^{\rm empirical}
    =\mu_0+c_\mu\mu_p+\mathcal{V}(\kappa,\alpha,v_d/v_s).
    \label{eq:empirical}
\end{equation}
Here,
$\mathcal{V}(\kappa,\alpha,v_d/v_s)\equiv
c_0\kappa\alpha^{1/2}(v_d/v_s)+c_1(v_d/v_s)^3$
is the sum of a damping-dependent term linear in $v_d/v_s$ and an empirical
nonlinear term cubic in $v_d/v_s$. An unweighted least-squares fit to
$\log\mu_b$ gives $\mu_0=2.3\times10^{-4}$, $c_\mu=2.0$, $c_0=0.33$, and
$c_1=0.72$.
Because the global fit combines data across system size, $\alpha$, $e$, and
shear rate, its values of $\mu_0$ and $c_\mu$ differ from those obtained from
the low-velocity fit at fixed system size and restitution coefficient in
Sec.~\ref{sec:friction}, but remain of the same order of magnitude.
Equation~\eqref{eq:empirical} provides an empirical
representation of the regime in which crystalline order is preserved and
single-dislocation glide persists.

\begin{figure}[!t]
    \centering
    \includegraphics[width=\linewidth]{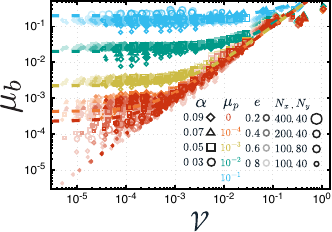}
    \caption{
    Empirical representation of the combined steady-state data:
    $\mu_b=-\bar\sigma_{xy}/\bar\sigma_{yy}$ against
    the combined rate variable $\mathcal{V}(\kappa,\alpha,v_d/v_s)$.
    The simulations cover
    $(N_x,N_y)=(100,40),(100,80),(200,40),(400,40)$ (symbol size),
    $\alpha=0.03,0.05,0.07,0.09$ (symbol shape),
    $\mu_p=0,10^{-4},10^{-3},10^{-2},0.1$ (color),
    $e=0.2,0.4,0.6,0.8$ (lighter shades for larger $e$), and 13 shear rates from
    $\dot\gamma=0.003$ to $3\,\mathrm{s}^{-1}$, comprising 4160 runs. The
    figure shows the 3589 points that satisfy
    $N(6-Z)\alpha^{3/2}\leq0.5$, excluding the
    regime in which the crystalline structure breaks down. For each $\mu_p$,
    the data for different $\alpha$, $e$, and system sizes collapse reasonably
    well. The dashed curves show the global fit
    $\mu_b=\mu_0+c_\mu\mu_p+\mathcal{V}(\kappa,\alpha,v_d/v_s)$
    [Eq.~\eqref{eq:empirical}] and agree reasonably well with the simulation
    data.}
    \label{fig:empirical_all}
\end{figure}

Figure~\ref{fig:empirical_all} compares this representation with the combined
steady-state data by plotting
$\mu_b=-\bar\sigma_{xy}/\bar\sigma_{yy}$ against
$\mathcal{V}(\kappa,\alpha,v_d/v_s)$. The data cover
$(N_x,N_y)=(100,40),(100,80),(200,40),(400,40)$,
$0.03\leq\alpha\leq0.09$, $0\leq\mu_p\leq0.1$, $0.2\leq e\leq0.8$, and
$0.003\leq\dot\gamma\leq3\,\mathrm{s}^{-1}$, comprising 4160 runs. For the
plot and global fit, we retain the 3589 runs satisfying
$N(6-Z)\alpha^{3/2}\leq0.5$, thereby excluding the regime in which the
crystalline structure breaks down. Outside the small-system,
$\alpha=0.03$, and $v_d/v_s\simeq1$ cases, variations in $\alpha$, $e$, and
system size fall close to a common trend at fixed $\mu_p$. The global
representation in Eq.~\eqref{eq:empirical}, shown by the dashed curves,
reproduces the simulation results reasonably well.

\section{Discussion}

\subsection{Comparison with other systems}

We first summarize the rheological distinction between conventional dense
amorphous granular materials and the present crystal containing a single
dislocation. In amorphous granular rheology, the shear-to-normal stress ratio
is commonly expressed as a function of the inertial number
$I=\dot\gamma d/\sqrt{\bar\sigma_{yy}/\rho}$, which compares the imposed
shear rate with the characteristic inertial rate
$\sqrt{\bar\sigma_{yy}/\rho}/d$ set by the particle size, density, and normal
stress~\cite{GDR-MiDi2004-ay,Andreotti2013-yo,Jop2006-yn,Da_Cruz2005-lr}.
In the conventional $\mu(I)$ picture, the stress ratio approaches a
quasistatic plateau as $I\to0$ and increases as $I$ approaches order unity.
The present crystal behaves differently. Its stress ratio varies
strongly within $I\simeq10^{-6}$--$10^{-3}\ll1$
[Fig.~\ref{fig:ss}(a)], and the data remain separated in the conventional
$\mu$--$I$ representation [Fig.~\ref{fig:mu-I}]. As in amorphous granular
rheology, the shear stress is normalized by the normal stress. Its rate
dependence, however, is organized more successfully by $v_d/v_s$, the
dislocation velocity normalized by the characteristic sound-speed scale
[Fig.~\ref{fig:stress_rate}(c)]. This difference reflects the fact that the
macroscopic deformation is carried by the glide of a single dislocation
through Orowan kinematics. At small $v_d/v_s$, the stress ratio is only
$10^{-4}$--$10^{-3}$ [Fig.~\ref{fig:stress_rate}(c)], far below the typical
$0.1$--$0.4$ range of dense amorphous granular flows
~\cite{Andreotti2013-yo}. This exceptionally small stress ratio accompanies deformation concentrated around a single gliding dislocation, whereas plastic flow in amorphous materials is carried by many localized rearrangements distributed in space and time~\cite{Nicolas2018-zj}.

The present granular crystal resembles atomic, colloidal, and plasma crystals
in that dislocation motion carries macroscopic plastic deformation
~\cite{Olmsted2005-ls,Fan2021-fs,Kim2024-wh,Nosenko2007-sd,Nosenko2011-yj}.
The mechanisms governing dislocation-mediated rheology, however, differ
among these systems. In atomic crystals, dislocations can
spontaneously nucleate and multiply during deformation, and interactions
within the resulting dislocation population give rise to collective
strengthening such as Taylor hardening~\cite{Fan2021-fs}. Moving
dislocations can also experience finite-temperature phonon drag
~\cite{Olmsted2005-ls,Fan2021-fs}. Dislocation interactions have also been
observed in colloidal crystals, whose particle dynamics are generally
overdamped by the suspending fluid~\cite{Kim2024-wh,Schall2004-eq}. In the
present athermal granular system, the low-velocity stress ratio instead
contains a single-dislocation lattice contribution analogous to the Peierls
resistance, together with a contribution from interparticle friction. At
intermediate $v_d/v_s$, contact damping, controlled through the restitution
coefficient, primarily governs the rate-dependent increase in stress. The
macroscopic rheology therefore directly reflects the frictional and
dissipative contact interactions characteristic of athermal granular matter.

\subsection{Explanation of the data}
\label{sec:explanation}

We now give a possible physical interpretation of the steady rheology. Even
without interparticle friction, the stress ratio approaches a finite plateau
at low shear rates [Fig.~\ref{fig:stress_rate}(c)]. We associate this plateau
with the quasistatic shear stress required for the dislocation to repeatedly
cross the periodic lattice barrier, releasing stored elastic energy during
each advance. As in amorphous granular materials, the normal stress sets the
relevant stress scale, so the low-rate shear stress can be written as
$-\bar\sigma_{xy}\simeq\mu_0\bar\sigma_{yy}$. The exceptionally small value
of $\mu_0$ reflects the localization of deformation around a single
dislocation [Fig.~\ref{fig:stress_rate}(c)].

At small $v_d/v_s$, interparticle friction provides an additional
contribution. Let $\zeta$ denote the extent of the dislocation region along
the glide direction. The dislocation crosses this region in a time
$\zeta/v_d$, during which approximately $\zeta/d$ particles each move by a
distance of order $d$ to a neighboring lattice position. Because the
friction force per particle scales as
$\mu_p\bar\sigma_{yy}d^2$, the frictional work per particle scales as
$\mu_p\bar\sigma_{yy}d^3$. Multiplying this by approximately $\zeta/d$
participating particles, the energy dissipated during this time scales as
$(\zeta/d)\mu_p\bar\sigma_{yy}d^3
=\mu_p\bar\sigma_{yy}d^2\zeta$. After subtracting the frictionless
contribution associated with elastic-energy release, the excess macroscopic
work is $(\mu_b-\mu_0)\bar\sigma_{yy}\dot\gamma LHd\,\zeta/v_d$.
Balancing these energies and using the Orowan relation
$\dot\gamma=bv_d/(LH)$ together with $b=(1-\alpha)d\simeq d$ gives
$\mu_b-\mu_0\sim\mu_p$. This explains the low-velocity
relation $\mu_b\simeq\mu_0+c_\mu\mu_p$ observed in
Fig.~\ref{fig:stress_rate_mu}(a).

The effect of the restitution coefficient at intermediate velocities can be
estimated from contact damping within the dislocation region
[Fig.~\ref{fig:damping}(a)]. The dislocation region contains approximately
$\zeta/d$ participating particles. As the region passes, each particle moves
by a distance of order $d$ over a time $\zeta/v_d$, giving a characteristic
velocity $d/(\zeta/v_d)=dv_d/\zeta$. This velocity also sets the relative
contact velocity between particles on opposite sides of the glide plane, so
that $v_{\rm rel}\sim dv_d/\zeta$. Because the damping force scales as
$\eta v_{\rm rel}$, the dissipation rate per contact scales as
$\eta v_{\rm rel}^2$ [Eq.~\eqref{eq:normal_force}], and the total damping
dissipation rate scales as
$(\zeta/d)\eta v_{\rm rel}^2\sim\eta(d/\zeta)v_d^2$. Across the low- and
intermediate-velocity regimes, the data show
$N(6-Z)\sim\alpha^{-3/2}$. If this deficit is proportional to the number of
particles in the dislocation region, then $N(6-Z)\sim\zeta/d$ and hence
$\zeta/d\sim\alpha^{-3/2}$, giving a damping dissipation rate of order
$\eta\alpha^{3/2}v_d^2$. An energy balance between this
dissipation rate and the damping-related part of the macroscopic input power
determines the contact-damping contribution to the stress ratio. Using the
Orowan relation [Eq.~\eqref{eq:orowan}],
$b=(1-\alpha)d\simeq d$, $\eta\sim\kappa\rho d^2v_s$ from
Eq.~\eqref{eq:damping_coefficient}, and
$\bar\sigma_{yy}\sim E\alpha^{3/2}$ gives a contact-damping contribution to
$\mu_b$ of order $\kappa\alpha^{1/2}v_d/v_s$ in the intermediate-velocity regime,
where contact damping dominates. Both the linear dependence on
$\kappa v_d/v_s$ and the factor $\alpha^{1/2}$ are consistent with the data
[Figs.~\ref{fig:damping}(a) and~\ref{fig:empirical_all}].

As $v_d/v_s$ approaches unity, the lattice has progressively less time to
relax between successive dislocation advances. The stress then develops a
nonlinear dependence on $v_d/v_s$, while contact damping becomes
comparatively less important. Over the velocity range studied, this
nonlinear increase is represented empirically by a $(v_d/v_s)^3$ term
[Figs.~\ref{fig:damping}(a) and~\ref{fig:empirical_all}]. At still larger
$v_d/v_s$, the coordination deficit rises sharply and the crystalline
structure breaks down [Figs.~\ref{fig:stress_rate}(d) and
\ref{fig:damping}(b)]. The deformation can then no longer be described as
single-dislocation glide.

\section{Conclusion}

We investigated the steady rheology of a granular crystal containing a
single dislocation. Our simulations show that
(i) within the single-dislocation regime, the steady shear-to-normal stress
ratio $\mu_b$ is not described by the conventional $\mu(I)$ rheology but is
reasonably organized as a function of $v_d/v_s$
[Figs.~\ref{fig:stress_rate}(c) and~\ref{fig:mu-I}];
(ii) interparticle friction primarily affects the low-$v_d/v_s$ regime,
where it raises the stress-ratio plateau
[Fig.~\ref{fig:stress_rate_mu}(a)];
(iii) contact damping, controlled by the restitution coefficient, primarily
governs the rate-dependent increase at intermediate $v_d/v_s$
[Fig.~\ref{fig:damping}(a)];
(iv) a pronounced nonlinear velocity dependence emerges as $v_d/v_s$
approaches unity
[Figs.~\ref{fig:damping}(a) and~\ref{fig:empirical_all}], while at still
larger $v_d/v_s$ crystalline order breaks down and the single-dislocation
description no longer applies
[Figs.~\ref{fig:stress_rate}(d) and~\ref{fig:damping}(b)]; and
(v) energy-balance arguments provide a qualitative account of the frictional
and damping contributions (Sec.~\ref{sec:explanation}).

These findings show that the steady rheology is governed by dislocation
dynamics coupled to the energy-dissipation mechanisms characteristic of
granular materials, namely interparticle friction and contact damping.
Interactions among multiple dislocations, pinning caused by point defects or
particle-size dispersity, and dislocation nucleation and annihilation provide
natural directions for extending the present single-defect rheology toward
defect-mediated collective plasticity in granular media.

\section*{Acknowledgments}

This work was supported by JSPS KAKENHI Grant Number JP25K17359. The
computation in this work was performed using the facilities of the
Supercomputer Center, the Institute for Solid State Physics, the University
of Tokyo (ISSPkyodo-SC-2026-Ba-0021).

\section*{Author Declarations}

\subsection*{Conflict of Interest}

The author has no conflicts to disclose.

\subsection*{Author Contributions}

Fumiaki Nakai: Conceptualization; Data curation; Formal analysis;
Funding acquisition; Investigation; Methodology; Project administration;
Resources; Software; Validation; Visualization; Writing -- original draft;
Writing -- review \& editing.

\section*{Data Availability}

The data and code that support the findings of this study are openly available
in Zenodo at \url{https://doi.org/10.5281/zenodo.22111447}~\cite{Nakai2026-zenodo}.

\FloatBarrier
\appendix
\section{Comparison with inertial-number scaling}

\begin{figure}[h]
    \centering
    \includegraphics[width=0.9\linewidth]{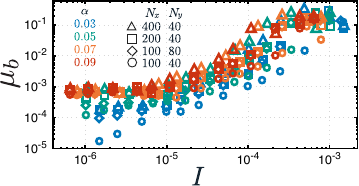}
    \caption{Stress ratio $\mu_b$ against the inertial number $I$ for the
    same frictionless data as Fig.~\ref{fig:stress_rate}: $\mu_p=0$, $e=0.4$,
    $\alpha=0.03,0.05,0.07,0.09$,
    $(N_x,N_y)=(100,40),(200,40),(100,80),(400,40)$, and all 13 rates from
    $\dot\gamma=0.003$ to $\SI{3}{\per\second}$. Each point is a final-20\%
    average. Colors denote $\alpha$, and open-symbol shapes denote
    $(N_x,N_y)$. The conventional variable $I$ leaves the curves separated.}
    \label{fig:mu-I}
\end{figure}

Figure~\ref{fig:mu-I} replots the data from
Fig.~\ref{fig:stress_rate} in the conventional $\mu$--$I$ representation.
The curves for different $\alpha$ and system sizes remain separated, showing
that this representation does not organize the steady response.

\bibliographystyle{aipnum4-2}
\bibliography{ref}

\end{document}